\documentclass[prd,reprint,aps,amsmath,amssymb,showpacs,nofootinbib
]{revtex4-1}
\usepackage{graphicx}
\usepackage{enumitem}
\usepackage{slashed}
\usepackage{url}
\usepackage[pdftex]{hyperref}
\usepackage[T1]{fontenc}
\usepackage{bm}
\usepackage{color}
\usepackage{bbm}

\begin{document}

\title{Cosmic neutrino background search experiments as decaying dark matter detectors}
\author{David McKeen}
\email{mckeen@triumf.ca}
\affiliation{TRIUMF, 4004 Wesbrook Mall, Vancouver, BC V6T 2A3, Canada}                          
\date{\today}

\begin{abstract}
We investigate the possibility that particles that are long-lived on cosmological scales, making up part or all of the dark matter, decay to neutrinos that have present-day energies around an eV. The neutrinos from these decays can potentially be visible at experiments that hope to directly observe the cosmic neutrino background through neutrino capture on tritium, such as PTOLEMY. In the context of a simple model that can realize such decays, we discuss the allowed signatures at a PTOLEMY-like experiment given current cosmological constraints.
\end{abstract}

\maketitle

\section{Introduction}
Particles with lifetimes on cosmological scales that decay to neutrinos arise in a number of new physics contexts, such as lepton number violation and the generation of neutrino masses~\cite{Chikashige:1980ui,*Schechter:1981cv,*Babu:1988ki}, and can even comprise some or all of the dark matter of the universe~\cite{Rothstein:1992rh,*Berezinsky:1993fm,*Lattanzi:2007ux,*Bazzocchi:2008fh,*Frigerio:2011in,*Queiroz:2014yna,*Wang:2016vfj,*Rojas:2017sih,*Brune:2018sab,Lattanzi:2013uza,Garcia-Cely:2017oco,*Heeck:2017kxw,Heeck:2018lkj}. The present-day energy of the neutrinos from these decays can span a wide range, making them accessible at, e.g., existing dark matter direct detection and neutrino experiments~\cite{PalomaresRuiz:2007ry,*Covi:2008jy,*Covi:2009xn,*Cui:2017ytb,Garcia-Cely:2017oco,*Heeck:2017kxw,Heeck:2018lkj}.

In this paper we will consider the decays of long-lived particles to neutrinos that carry very little energy today, in the neighborhood of an eV. These neutrinos are slightly more energetic than the standard cosmic neutrino background (C$\nu$B)---the thermal relic neutrinos that decoupled from the plasma when the universe had a temperature of a few MeV. The temperature of the C$\nu$B is $\sim10^{-4}~\rm eV$ today, less than the scale of the atmospheric and solar neutrino mass splittings, and therefore at least two mass eigenstates in the C$\nu$B are now nonrelativistic; the most energetic of these neutrinos have energy $E_{\rm C\nu B}\simeq m_\nu\gtrsim 0.05~\rm eV$. There is an upper limit on the neutrino masses, hence on the energy of the C$\nu$B, from observations of the cosmic microwave background (CMB) and baryon acoustic oscillations of $\sum m_\nu<0.12~\rm eV$ at 95\%~confidence level~\cite{Aghanim:2018eyx}. Terrestrial experiments, in comparison, limit the neutrino mass at 95\%~confidence level to less than $2.8~\rm eV$~\cite{Weinheimer:1999tn,*Lobashev:1999tp}.

The present-day number density of C$\nu$B neutrinos is very large, with a cosmic average $n_{\rm C\nu B}\simeq330~\rm cm^{-3}$. Despite this enormous density, scattering of C$\nu$B neutrinos is highly suppressed since their energies are minuscule and they couple to matter only via the weak interaction. Furthermore, the tiny energies involved make distinguishing from backgrounds very troublesome. Both of these facts mean that detecting the C$\nu$B is extremely challenging. The most promising technique to detect the very low energy C$\nu$B---neutrino capture on $\beta$-decaying nuclei, $\nu+(A,Z)\to e^-+(A,Z+1)$, where a nucleus, $(A,Z)$, is labeled by its mass number, $A$, and atomic number, $Z$---was first proposed by Weinberg in 1962~\cite{Weinberg:1962zza} and more recently studied in~\cite{Cocco:2007za,*Lazauskas:2007da,*Blennow:2008fh,*Cocco:2009rh,*Li:2010sn,*Faessler:2011qj,Long:2014zva}. This process benefits from the lack of a threshold energy and its signature is the production of an electron with energy above the end point of natural $\beta$ decay, $(A,Z)\to \bar\nu+e^-+(A,Z+1)$. For a neutrino of mass $m_\nu$ and energy $E_\nu$, the shift above the end point is $\sim m_\nu+E_\nu$. In the case of nonrelativistic C$\nu$B neutrinos, this shift is roughly $2m_\nu$. Of course, extremely good resolution on the $e^-$ energy is required to resolve this gap given the relatively large rate of $\beta$ decay compared to capture of C$\nu$B neutrinos.

Recently, the PTOLEMY experiment has proposed~\cite{Betts:2013uya,*Baracchini:2018wwj} to tackle these difficulties using a target of 100~g of tritium (i.e. $A=3$, $Z=1$ in the expressions above) implanted on a graphene substrate along with MAC-E filtering, radio frequency monitoring, and advanced calorimetry to measure the $e^-$ energy. Tritium has a relatively small rate for $\beta$ decay, $\Gamma_\beta\simeq(17~\rm yr)^{-1}$ and implanting it on graphene serves to reduce intrinsic broadening of the $e^-$ energy from molecular effects. This could potentially allow for energy resolution as small as $0.1~\rm eV$, which is needed to successfully probe the C$\nu$B.

This paper is organized as follows. In Sec.~\ref{sec:densitiesrates}, we describe a simple model containing a long-lived particle that decays to neutrinos, and discuss the number density and energy distribution of neutrinos  allowed by cosmological observations. We see that decays that take place after recombination, when photons decouple, are the most promising to be probed at experiments sensitive to eV-scale neutrinos, such as PTOLEMY. We discuss the signature of this scenario at a PTOLEMY-like experiment in detail in Sec.~\ref{sec:sig}, considering both the diffuse and local contributions to the flux, and compare the reach to that of cosmological observations. In Sec.~\ref{sec:conc}, we conclude.

\section{Densities and distributions}
\label{sec:densitiesrates}
The simple model we consider involves only a long-lived, real scalar particle $J$ with mass $m_J$. It couples to the light neutrinos $\nu$ through an interaction of the form
\begin{equation}
{\cal L}_{\rm int}=-\frac{g}{2}J\nu\nu+{\rm H.c.}
\label{eq:JLag}
\end{equation}
In this expression, $\nu$ is a two-component, left-chiral spinor field, and the interaction is gauge invariant if, e.g., it comes from a sterile admixture of the light neutrinos. This coupling leads to the decay $J\to\nu\nu$. Assuming that this is the only $J$ decay mode at tree level, its lifetime is
\begin{equation}
\tau_J=\frac{32\pi}{|g|^2m_J}=2\times10^9~{\rm yr}\left|\frac{10^{-15}}{g}\right|^2\left(\frac{\rm eV}{m_J}\right).
\end{equation}
Here and in what follows, we take the neutrino masses to be negligible compared to $m_J$. A natural candidate for $J$ is the Majoron associated with the spontaneous breaking of lepton number at a scale $f$. In this case, the coupling is $g=i m_\nu/f$ which could easily be tiny for $f$ above the TeV scale. We have normalized the $J$ mass on an eV for later convenience. For now, we are agnostic about the flavor structure of the couplings in Eq.~(\ref{eq:JLag}) but will return to this point in Sec.~\ref{sec:sig}.

We assume that the $J$'s are produced nonthermally and are nonrelativistic at cosmologically interesting times.\footnote{For instance, production could proceed through misalignment as in the case of the axion, which motivates considering $J$ as the Nambu-Goldstone boson of some approximate global symmetry. Studying this or other production mechanisms in detail, as well as the generation of $m_J$,  is outside the scope of this paper and left for future work.} We can then simply write down their number density,
\begin{equation}
\begin{aligned}
n_J(t)&=\Omega_J\frac{\rho_{{\rm cr},0}}{m_J}\left[\frac{a(t_0)}{a(t)}\right]^3e^{-t/\tau_J}
\\
&=\frac{63}{\rm cm^3}\left(\frac{\Omega_J/\Omega_{\rm dm}}{0.05}\right)\left(\frac{\rm eV}{m_J}\right) \left[\frac{a(t_0)}{a(t)}\right]^3e^{-t/\tau_J}.
\end{aligned}
\label{eq:nJ}
\end{equation}
In this expression, $a(t)$ is the scale factor of the universe, $t_0=13.8~\rm Gyr$ is its age, $\Omega_J$ is the $J$ energy density in units of the critical energy density of the universe today, $\rho_{{\rm cr},0}=10.5\,h^2~{\rm keV/cm^3}$, $\Omega_{\rm dm}=0.12\,h^{-2}$ is that of dark matter, and $h\simeq0.68$ is the Hubble constant today in units of $100~\rm km/s/Mpc$. $\Omega_J/\Omega_{\rm dm}$ is the fraction of dark matter that $J$ particles would comprise today if they did not decay, which we normalize here to $5\%$. In what follows, we will suppress the argument of the scale factor and set its value today to unity, $a_0=a(t_0)=1$. We will also require that the $J$'s and their decay products do not grossly disturb the evolution of the universe from the standard cosmology, with radiation domination at redshifts $z=a^{-1}-1\gtrsim z_{\rm eq}\simeq3300$, followed by matter domination, then more recently vacuum energy domination.

The number density of neutrinos produced in $J$ decays simply follows from Eq.~(\ref{eq:nJ}),
\begin{equation}
\begin{aligned}
\tilde n_\nu(t)&=\frac{2\,\Omega_J\rho_{{\rm cr},0}}{m_J}\frac{1-e^{-t/\tau_J}}{a^3}
\\
&=\frac{130}{\rm cm^3}\left(\frac{\Omega_J/\Omega_{\rm dm}}{0.05}\right)\left(\frac{\rm eV}{m_J}\right)\frac{1-e^{-t/\tau_J}}{a^3},
\label{eq:nnu}
\end{aligned}
\end{equation}
where we use a tilde to distinguish this population from the standard neutrinos.

There are essentially three qualitatively different regimes for the $J$ lifetime in terms of its cosmological effects: (i) before neutrino decoupling, $\tau_J< t_{\rm dec}\simeq 0.2~\rm s$, (ii) after neutrino decoupling and before recombination, $t_{\rm dec}<\tau_J<t_{\rm rec}\sim 4\times10^5~{\rm yr}$, and (iii) after recombination, $\tau_J>t_{\rm rec}$. Case (i) is not observable since the neutrinos simply thermalize with the plasma. Cases (ii) and (iii) can potentially lead to a nonstandard population of neutrinos today. Crucially, cases (ii) and (iii) affect the observation of the CMB in different ways which we discuss below.

\subsection{Decays before recombination ($\tau_J<t_{\rm rec}$)}
\label{sec:ndensityRD}
In this case, the neutrinos from $J$ decays contribute to the energy density in relativistic species at early times. This is constrained by the observation of the CMB which is record of the universe at around $t_{\rm rec}$.

The extra contribution from $J$ decays can be conveniently parametrized by a shift of the effective number of relativistic degrees of freedom, $\Delta N_{\rm eff}$, with 
\begin{equation}
\Delta N_{\rm eff}=\frac{8}{7}\left(\frac{11}{4}\right)^{4/3}\frac{\tilde\rho_\nu}{\rho_\gamma},
\label{eq:Neff_def}
\end{equation}
where $\rho_\gamma$ and $\tilde\rho_\nu$ are the energy densities in photons and neutrinos from $J$ decay, respectively. A nonzero $\Delta N_{\rm eff}$ can affect the CMB by changing the expansion rate around the time of last scattering from its standard value. The current $95\%$~confidence level upper limit on $\Delta N_{\rm eff}$ from CMB and large-scale structure observations is $\Delta N_{\rm eff}<0.28$~\cite{Aghanim:2018eyx}. This upper limit can constrain the number of neutrinos from $J$ decay for $\tau_J<t_{\rm rec}$.

To determine this limit, we relate the energy density in neutrinos from $J$ decay to the $J$ number density,
\begin{equation}
\frac{d\tilde\rho_\nu}{dt}+4H\tilde\rho_\nu=\frac{m_J n_J}{\tau_J}.
\end{equation}
Using the fact that photons redshift like radiation, $\rho_\gamma\propto a^{-4}$, we can re-express this equation as
\begin{equation}
\frac{d(\tilde\rho_\nu/\rho_\gamma)}{dt}=\frac{m_J n_J}{\rho_\gamma\tau_J}=\frac{\Omega_J}{\Omega_\gamma}\frac{a}{\tau_J}e^{-t/\tau_J},
\label{eq:rhonu_diffeq}
\end{equation}
where $\Omega_\gamma=2.47\times10^{-5}\,h^{-2}$ is the present-day contribution of the CMB to the critical density. Assuming that the $J$'s do not come to dominate the energy budget of the universe at this time, the energy density is dominated by radiation and the scale factor depends on the time as $t=t_2a^2$ with $t_2=7.6\times10^{11}~{\rm yr}$. We can then integrate Eq.~\ref{eq:rhonu_diffeq} to find
\begin{equation}
\begin{aligned}
\frac{\tilde\rho_\nu}{\rho_\gamma}\Big|_{t\gg\tau_J}&=\frac{\sqrt\pi}{2}\frac{\Omega_J}{\Omega_\gamma} \sqrt{\frac{\tau_J}{t_2}}
\\
&=0.15\left(\frac{\Omega_J}{\Omega_{\rm dm}}\right)\sqrt{\frac{\tau_J}{10^3~\rm yr}}.
\end{aligned}
\label{eq:rhoratio}
\end{equation}
Using Eqs.~(\ref{eq:nnu}), (\ref{eq:Neff_def}), and (\ref{eq:rhoratio}), we can re-express $\tilde n_\nu$ today in terms of $\Delta N_{\rm eff}$,
\begin{equation}
\begin{aligned}
\tilde n_\nu(t_0)&=\frac{1.0\times10^{3}}{\rm cm^3}\left(\frac{\Delta N_{\rm eff}}{0.28}\right)\left(\frac{\rm eV}{m_J}\right)\sqrt{\frac{10^3~\rm yr}{\tau_J}}.
\label{eq:nnuRD}
\end{aligned}
\end{equation}

An additional constraint comes from the fact that $J$ particles redshift like matter before their decay so that their energy density can come to exceed that in radiation, causing an early period of matter domination. This would conflict with the usual picture of radiation domination from primordial nucleosynthesis until matter-radiation equality at $t_{\rm eq}\sim t_{\rm rec}$; we can use this to limit $\Omega_J$ for $\tau_J\lesssim t_{\rm eq}\sim t_{\rm rec}$. The ratio of the energy density in $J$'s to that in radiation during this era is
\begin{equation}
\begin{aligned}
\frac{\rho_J}{\rho_r}&=\frac{2}{g_\ast}\frac{m_J n_J}{\rho_\gamma}=\frac{2}{g_\ast}\frac{\Omega_J}{\Omega_\gamma}a\, e^{-t/\tau_J}
\\
&=\frac{2}{g_\ast}\frac{\Omega_J}{\Omega_\gamma}\sqrt{\frac{t}{t_2}}e^{-t/\tau_J},
\end{aligned}
\end{equation}
with $g_\ast=3.36$. Requiring that this ratio is less than unity gives the constraint
\begin{equation}
\frac{\Omega_J}{\Omega_{\rm dm}}<23\sqrt{\frac{10^3~\rm yr}{\tau_J}}.
\end{equation}
Note that this is equivalent to a rather weak limit on the number of relativistic degrees of freedom from CMB observations of $\Delta N_{\rm eff}<15$. However, it can become an important constraint in a scenario where only a fraction of $m_J$ ($\sim 0.28/15\simeq 2\%$ or less) goes into relativistic degrees of freedom in each $J$ decay. This could be the case, e.g., in neutrino portal dark matter models~\cite{Bertoni:2014mva,*Gonzalez-Macias:2016vxy,*Ibarra:2016fco,*Escudero:2016ksa,*Batell:2017rol,*Batell:2017cmf,*Schmaltz:2017oov} with a small splitting in the dark sector between a fermion, $\chi$ and a scalar $\phi$. These interact with neutrinos via the effective operator $\phi\bar\chi HL/\Lambda\to(v/\Lambda)\phi\bar\chi\nu$ where $H$ and $L$ are the Higgs and lepton doublets respectively. The heavier state in the dark sector, $\phi$ for instance, could be long-lived and decay through this operator, $\phi\to\chi\nu$ where the neutrino has energy $\sim m_\phi-m_\chi$ in the $\phi$ rest frame.

Lastly, looking at Eq.~(\ref{eq:nnuRD}),we might naively think that we can arrange for a number density of neutrinos from $J$ decays before recombination that is much larger than that of the C$\nu$B neutrinos. This could be the case if the $J$'s were very light and cold so that their energy density is suppressed while their number density is large---this relies on the $J$'s remaining unthermalized with the C$\nu$B neutrinos. This requirement can be used to set an upper limit on the strength of the $J$-$\nu$ interaction or, equivalently, a lower limit on $\tau_J$. Production of $J$'s through $\nu\nu\to J$ happens most readily at $T_\nu\sim m_J$ and the rate for this is roughly $g^2T_\nu\sim1/\tau_J$. To keep the $J$'s out of equilibrium, we require that this is less than the Hubble rate at $T_\nu\sim m_J$ which is $\sim m_J^2/M_{\rm Pl}$ with $M_{\rm Pl}\simeq 10^{19}~\rm GeV$ the Planck mass. This implies that $g^2\lesssim m_J/M_{\rm Pl}$ or $\tau_J\gtrsim 10^5~{\rm yr}\,({\rm eV}/m_J)^2$. Using this in Eq.~(\ref{eq:nnuRD}) one obtains an upper bound on the present-day density of neutrinos from $J$ decays roughly comparable to that in the C$\nu$B, ${\cal O}(100~\rm cm^{-3})$. The energy distribution of such neutrinos today would be indistinguishable from that of the C$\nu$B; i.e., they would also be nonrelativistic. 

Similar situations where nonthermally produced neutrinos, such as the right-chiral component of light Dirac neutrinos, evade constraints on light degrees of freedom and lead to an enhancement of the C$\nu$B signal have already been explored in, e.g.~\cite{Chen:2015dka,*Zhang:2015wua}. Dark matter that comes into thermal equilibrium with neutrinos after neutrino decoupling but before recombination has been studied extensively in~\cite{Berlin:2017ftj,*Berlin:2018ztp}.

\subsection{Decays after recombination ($\tau_J>t_{\rm rec}$)}
Nonrelativistic $J$'s that decay to neutrinos after recombination act as a decaying component of the dark matter. This alters the expansion history of the universe between last scattering and today which can change the precise pattern of CMB angular anisotropies as well as the growth of structure. These effects have been analyzed in detail in Refs.~\cite{Lattanzi:2013uza,Poulin:2016nat,Chudaykin:2016yfk,*Chudaykin:2017ptd} and even proposed as an explanation of tensions in cosmological data~\cite{Berezhiani:2015yta,*Enqvist:2015ara}. This limits $\tau_J$ and the energy density in $J$'s and, consequently, the number density of neutrinos produced in $J$ decay.

For lifetimes short compared to the age of the universe (but long compared to $t_{\rm rec}$), Ref.~\cite{Poulin:2016nat} obtains a bound using CMB observables on the $J$ energy density of $\Omega_J/\Omega_{\rm dm}<0.038$ at 95\%~confidence level. For longer lifetimes, the constraint is roughly $\Omega_J/\Omega_{\rm dm}<0.09\,(\tau_J/15~\rm Gyr)$. Using these constraints in Eq.~(\ref{eq:nnu}), the current neutrino number density is then limited to be
\begin{align}
&\tilde n_\nu(t_0)\lesssim\frac{95}{\rm cm^3}\left(\frac{\rm eV}{m_J}\right)
\label{eq:nnuMD}
\\
&\times
\begin{cases}
1, &\tau_J\lesssim 12~{\rm Gyr}\\
2.3\left(1-t_0/2\tau_J\right) ,&12~{\rm Gyr}\lesssim \tau_J\lesssim 160~{\rm Gyr}\\
2.3\left(160~{\rm Gyr}/\tau_J\right),&\tau_J\gtrsim 160~{\rm Gyr}.
\end{cases}
\nonumber
\end{align}

We observe that neutrino number densities comparable to that in the C$\nu$B from $J$ decay are possible for $\tau_J$ larger or smaller than $t_{\rm rec}$ without being ruled out by cosmological data. We now turn to the question of their energy distribution which greatly impacts their detectability.

\subsection{Energy spectrum}
In this section, we find the (cosmic average) energy density of the neutrinos produced in $J$ decays, assuming $m_\nu\ll m_J$. At the time of their production, each neutrino has energy $m_J/2$. Afterwards, the energy redshifts with the expansion of the universe so that today it is $E_\nu=(m_J/2)a$ where $a$ is the scale factor at the time of decay ($a_0=1$). Since the $J$'s do not decay all at once, the neutrino spectrum today is not monochromatic, broadened by the expansion of the universe over the course of the $J$ decays. More quantitatively, the spectrum today is simply related to the change in the comoving number density of neutrinos with scale factor,
\begin{equation}
\begin{aligned}
\frac{d\tilde n_\nu}{dE_\nu}&\equiv\frac{d(a^3\tilde n_\nu)}{dE_\nu}\Big|_{a= 2E_\nu/m_J}=\frac{2}{m_J}\frac{d(a^3\tilde n_\nu)}{da}
\\
&=\frac{2\,\Omega_J\rho_{{\rm cr},0}}{m_J}\frac{e^{-t/\tau_J}}{H \tau_J E_\nu},
\end{aligned}
\label{eq:spectrum}
\end{equation}
where the time and Hubble rate are evaluated at $a=2E_\nu/m_J$. Assuming that during $J$ decay the Hubble rate depends on the scale factor simply as $H=(nt_n a^n)^{-1}$ ($n=2,3/2$ correspond to radiation and matter domination respectively), we can find the ``average'' energy of the neutrinos today, at which $d\tilde n_\nu/dE_\nu$ is maximized,
\begin{equation}
\begin{aligned}
E_\nu^{\rm av}=\frac{m_J}{2}\left[\frac{n-1}{n}\frac{\tau_J}{t_n}\right]^{1/n}.
\end{aligned}
\end{equation}
$t_n$ is a constant that depends on whether the universe was radiation or matter dominated at the time of decay. Again, assuming the $J$'s and their neutrino decay products do not change the evolution of the universe from the standard picture, $t_2$ is given above Eq.~(\ref{eq:rhoratio}) and $t_{3/2}=1.7\times10^{10}~{\rm yr}$ which apply to $\tau_J<t_{\rm eq}$ and $\tau_J>t_{\rm eq}$, respectively, where $t_{\rm eq}\sim t_{\rm rec}$ is the time of matter-radiation equality. Using these expressions the characteristic energy of the neutrinos today is
\begin{align}
\label{eq:Eav}
\frac{E_\nu^{\rm av}}{m_J}&=
\begin{cases}
1\times10^{-5}\sqrt{\tau_J/10^3~{\rm yr}}, &\tau_J\lesssim t_{\rm rec}\\
{\rm min}[0.04\left(\tau_J/{\rm Gyr}\right)^{2/3},\,0.5],&\tau_J\gtrsim t_{\rm rec},
\end{cases}
\end{align}
which ignores a slight correction due to the relatively recent transition to vacuum energy domination. Therefore, obtaining relativistic neutrinos today with $E_\nu\sim\rm eV$ requires $m_J\gtrsim 10~{\rm keV}$ ($m_J\lesssim 10~{\rm keV}$) for $\tau_J\lesssim t_{\rm rec}$ ($\tau_J\gtrsim t_{\rm rec}$), as we would expect given the redshift of matter-radiation equality, $z_{\rm eq}\simeq3300$. 

\begin{figure}
\begin{center}
\includegraphics[width=1.0\linewidth]{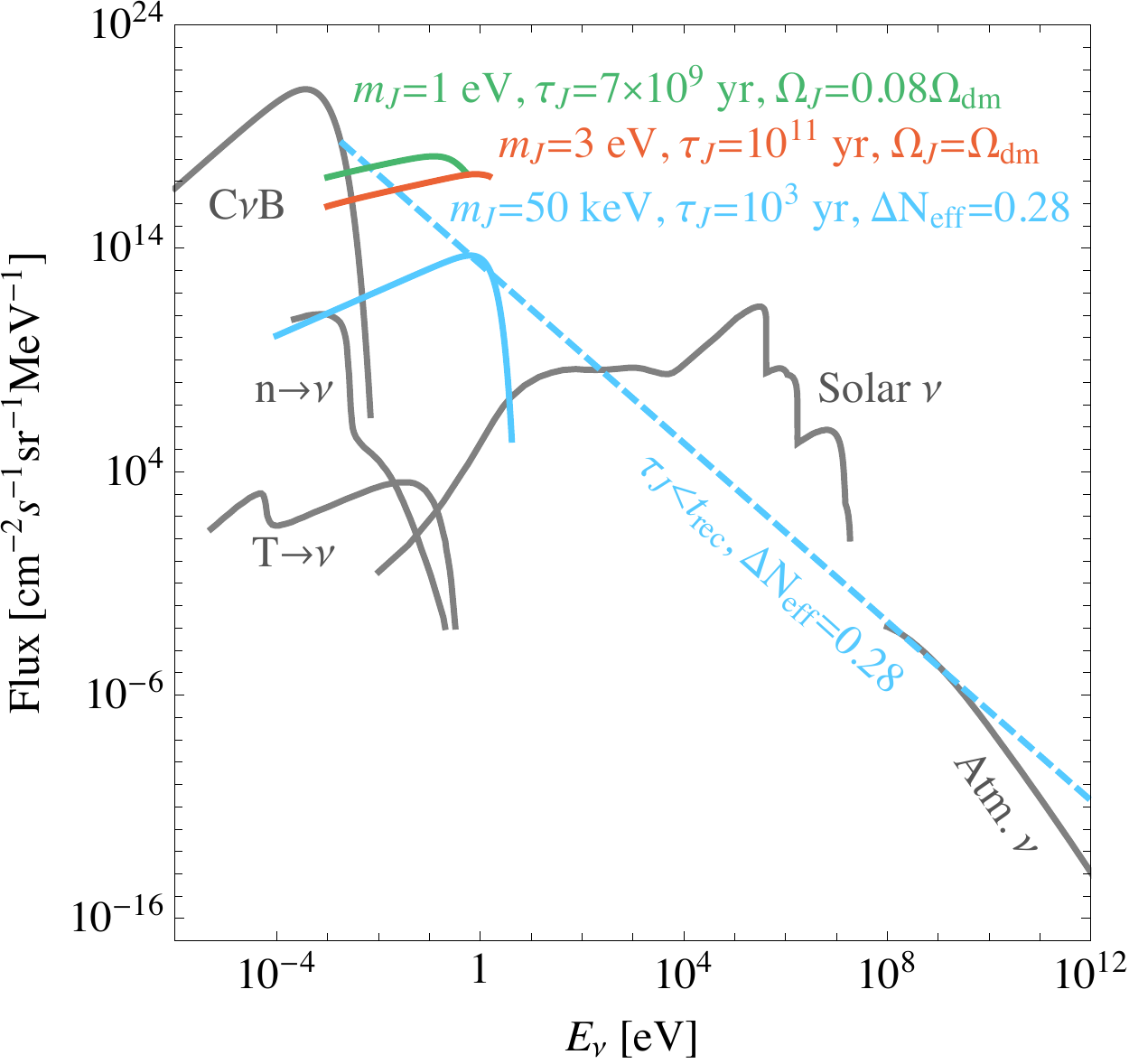}
\end{center}
\caption{The flux of neutrinos from $J\to\nu\nu$, averaged over the universe, along with C$\nu$B, solar $\nu$~\cite{Serenelli:2016dgz,Vitagliano:2017odj}, and atmospheric $\nu$~\cite{Honda:2011nf} neutrinos, as well as antineutrinos from the decay of neutrons and tritons produced in primordial nucleosynthesis~\cite{Ivanchik:2018fxy}. We show the flux for $m_J=50~\rm keV$, $\tau_J=10^3~\rm yr$, and, since the decays take place before recombination, $\Omega_J$ such that this gives $\Delta N_{\rm eff}=0.28$ (solid, cyan). In addition, we show the upper limit on any flux from a decay taking place before recombination from the upper limit $\Delta N_{\rm eff}=0.28$ (dashed, cyan). Also shown are the fluxes for two benchmark points for decays after $t_{\rm rec}$ in Sec.~\ref{sec:sig}: $m_J=1~\rm eV$, $\tau_J=7\times 10^{9}~\rm yr$, $\Omega_J=0.08\,\Omega_{\rm dm}$ (solid, green) and $m_J=3~\rm eV$, $\tau_J=10^{11}~\rm yr$, $\Omega_J=\Omega_{\rm dm}$ (solid, red).}\label{fig:flux}
\end{figure}

In Fig.~\ref{fig:flux}, we show the cosmic average flux of nonstandard neutrinos as functions of their energy today for $m_J=50~\rm keV$, $\tau_J=10^3~{\rm yr}<t_{\rm rec}$ and $\Omega_J$ corresponding to $\Delta N_{\rm eff}=0.28$ (solid cyan); $m_J=1~\rm eV$, $\tau_J=7\times10^9~\rm yr$, $\Omega_J/\Omega_{\rm dm}=0.08$ (solid, green); and $m_J=3~\rm eV$, $\tau_J=10^{11}~\rm yr$, $\Omega_J=\Omega_{\rm dm}$ (solid, red). We also show the upper limit on the flux from decays prior to recombination such that they contribute $\Delta N_{\rm eff}=0.28$ (dashed, cyan). To generate these fluxes we used the energy spectrum from Eq.~(\ref{eq:spectrum}) with the Hubble rate calculated for a universe with contributions to the energy density from baryons, $\Omega_{\rm b}=0.05$; dark energy, $\Omega_\Lambda=0.69$; dark matter, $\Omega_{\rm dm}=0.26$; and (standard) radiation, $\Omega_{r}=8.7\times10^{-5}$. For $\tau_J>t_{\rm rec}$, the dark matter is split up into a decaying component, $\Omega_J$, and a nondecaying component, $\Omega_{\rm dm}-\Omega_J$.

For comparison, the fluxes of C$\nu$B, solar $\nu$~\cite{Serenelli:2016dgz}, and atmospheric $\nu$~\cite{Honda:2011nf} neutrinos are plotted in Fig.~\ref{fig:flux} as well. We also show the results of the recent computations of the low energy thermal component of solar neutrinos~\cite{Vitagliano:2017odj} and the antineutrino flux from the decay of neutrons and tritons produced in primordial nucleosynthesis~\cite{Ivanchik:2018fxy}, which extend into the $0.1$-$1~\rm eV$ range but in relatively low numbers.

Because the number density is suppressed with increasing $m_J$, and cosmological limits are strong for small $\tau_J$, the most viable scenario to produce an observable number of neutrinos with $E_\nu\sim\rm eV$ today involves relatively light $J$'s, of order a few eV, with lifetimes comparable to the age of the universe. If the light neutrinos are massive, it is possible for early decays, before $t_{\rm rec}$, of light $J$'s to lead to a population of nonrelativistic neutrinos with present-day energy $E_\nu\sim m_\nu$. The signal of these neutrinos is indistinguishable from that of the C$\nu$B in such a case. Since, as seen in Sec.~\ref{sec:ndensityRD}, such neutrinos' number density can be at most comparable to the C$\nu$B, they can only lead to an ${\cal O}(1)$ increase in the event rate of massive C$\nu$B-like neutrinos, similar to scenarios explored in~\cite{Chen:2015dka,*Zhang:2015wua}. The standard C$\nu$B rate can vary by an ${\cal O}(1)$ factor depending on whether the neutrinos are Majorana or Dirac~\cite{Long:2014zva} as well as gravitational clustering~\cite{Ringwald:2004np} and focusing by the Sun~\cite{Safdi:2014rza}. For these reasons, in what follows we focus on decays of a light $J$ after recombination. Below, we examine the experimental signature of neutrinos from these decays at experiments searching for the C$\nu$B such as PTOLEMY.

\section{Observing the neutrinos from dark matter decay}
\label{sec:sig}
\subsection{Neutrino capture from diffuse $J\to\nu\nu$}
We are interested in the detection prospects for neutrinos with $E_\nu\sim\rm eV$ today which could show up in experiments searching for direct evidence of the C$\nu$B. The most promising technique to detect these neutrinos is neutrino capture on $\beta$-decaying nuclei. As a benchmark detector setup, we focus on the recently proposed experiment PTOLEMY~\cite{Betts:2013uya,*Baracchini:2018wwj}, which hopes to use a tritium target of mass $M_T=100~\rm g$ and to achieve an energy resolution of about $0.1~\rm eV$ (full width at half maximum) on recoiling electrons near the end point. This could probe neutrinos with energies as low as $0.1~\rm eV$.

The cross section for an electron neutrino with $E_\nu\ll{\rm keV}$ to capture on tritium is~\cite{Long:2014zva}
\begin{equation}
\sigma=3.83\times10^{-45}~{\rm cm}^2.
\label{eq:capturexsec}
\end{equation}
Using this with Eq.~(\ref{eq:nnu}) leads to a capture rate on tritium for neutrinos from $J$ decays, averaged over the universe, of
\begin{equation}
\begin{aligned}
R_{\rm cos}&=\frac{4.52}{\rm yr}\left(\frac{M_T}{100~\rm g}\right)\left(\frac{f_{\nu_e}}{1/2}\right)\left(\frac{\Omega_J/\Omega_{\rm dm}}{0.05}\right)
\\
&\quad\times\left(\frac{\rm eV}{m_J}\right)\left(1-e^{-t_0/\tau_J}\right).
\label{eq:rate}
\end{aligned}
\end{equation}
In this expression $f_{\nu_e}$ is the fraction of these neutrinos that are of electron flavor. While we have thus far not specified the flavor content of the neutrinos that $J$ couples to, we will discuss this briefly here.

Note that if $J$ is indeed the Majoron~\cite{Chikashige:1980ui,*Schechter:1981cv,*Babu:1988ki} it couples to the neutrino mass eigenstates with strength proportional to their masses, and thus the flavor content of these neutrinos does not oscillate~\cite{Garcia-Cely:2017oco,*Heeck:2017kxw}. First, consider the case that the neutrinos are light and not degenerate, $\sum m_\nu\lesssim0.1~\rm eV$. If the mass hierarchy is normal then $J\to\nu_3\nu_3$ is the dominant mode and $f_{\nu_e}\simeq0.03$. If instead it is inverted, $\Gamma_{J\to\nu_1\nu_1}\simeq\Gamma_{J\to\nu_2\nu_2}$ and $f_{\nu_e}\simeq1/2$. If the neutrinos are relatively heavy and degenerate with $\sum m_\nu\gtrsim0.1~\rm eV$ the rate into all mass eigenstates is comparable and $f_{\nu_e}\simeq1/3$. Different scenarios where $J$ couples to states that are not mass eigenstates lead to flavor oscillations and can give different values of $f_{\nu_e}$.

Since we are focusing on $\tau_J>t_{\rm rec}$, we can use the same limit on $\Omega_J$ as a function of $\tau_J$ from~\cite{Poulin:2016nat}  that lead to the upper limit on $\tilde n_\nu$ in Eq.~(\ref{eq:nnuMD}) to limit their capture rate. Doing so gives
\begin{align}
&R_{\rm cos}\lesssim\frac{3.44}{\rm yr}\left(\frac{M_T}{100~\rm g}\right)\left(\frac{f_{\nu_e}}{1/2}\right)\left(\frac{\rm eV}{m_J}\right)
\\&\times
\begin{cases}
1, &\tau_J\lesssim 12~{\rm Gyr}\\
2.3\left(1-t_0/2\tau_J\right) ,&12~{\rm Gyr}\lesssim \tau_J\lesssim 160~{\rm Gyr}\\
2.3\left(160~{\rm Gyr}/\tau_J\right),&\tau_J\gtrsim 160~{\rm Gyr},
\end{cases}
\nonumber
\end{align}
which, as expected, is comparable to the rates expected from the C$\nu$B. However, as we see from Eq.~(\ref{eq:Eav}), the neutrinos' energies can be around an eV today {\it without} requiring the neutrino masses to be large.

\subsection{Dark matter decay in the Galaxy}
Thus far, we have discussed only the contribution to a neutrino capture signal from diffuse $J$ decays averaged over the entire universe. Since the $J$'s act as dark matter today if $\tau_J$ is comparable to or larger than the age of the universe, their local density in the Milky Way can be greatly enhanced over the cosmic average. Using a Navarro-Frenck-White dark matter profile~\cite{Navarro:1996gj}, $\rho(r)\propto r^{-1}(1+r/r_s)^{-2}$ with scale radius $r_s=24~\rm kpc$ and local dark matter density of $0.3~\rm GeV/cm^3$ along with the capture cross section on tritium in Eq.~(\ref{eq:capturexsec}), we find a rate for capture from neutrinos produced in the Milky Way of
\begin{equation}
\begin{aligned}
R_{\rm MW}&=\frac{5.82}{\rm yr}\left(\frac{M_T}{100~\rm g}\right)\left(\frac{f_{\nu_e}}{1/2}\right)\left(\frac{\rm eV}{m_J}\right)
\\
&\quad\times\left(\frac{10~\rm Gyr}{\tau_J}\right)\left(\frac{\Omega_J/\Omega_{\rm dm}}{0.05}\right)e^{-t_0/\tau_J}.
\end{aligned}
\end{equation}
This signal is essentially monochromatic with $E_\nu=m_J/2$. For $\tau_J\lesssim5~\rm Gyr$, this rate is negligible but becomes important as $\tau_J$ is increased, so that for $\tau_J\gtrsim 100~\rm Gyr$ it is roughly comparable to that from cosmically averaged $J$ decays, $R_{\rm cos}$.

\begin{figure}
\begin{center}
\includegraphics[width=1.0\linewidth]{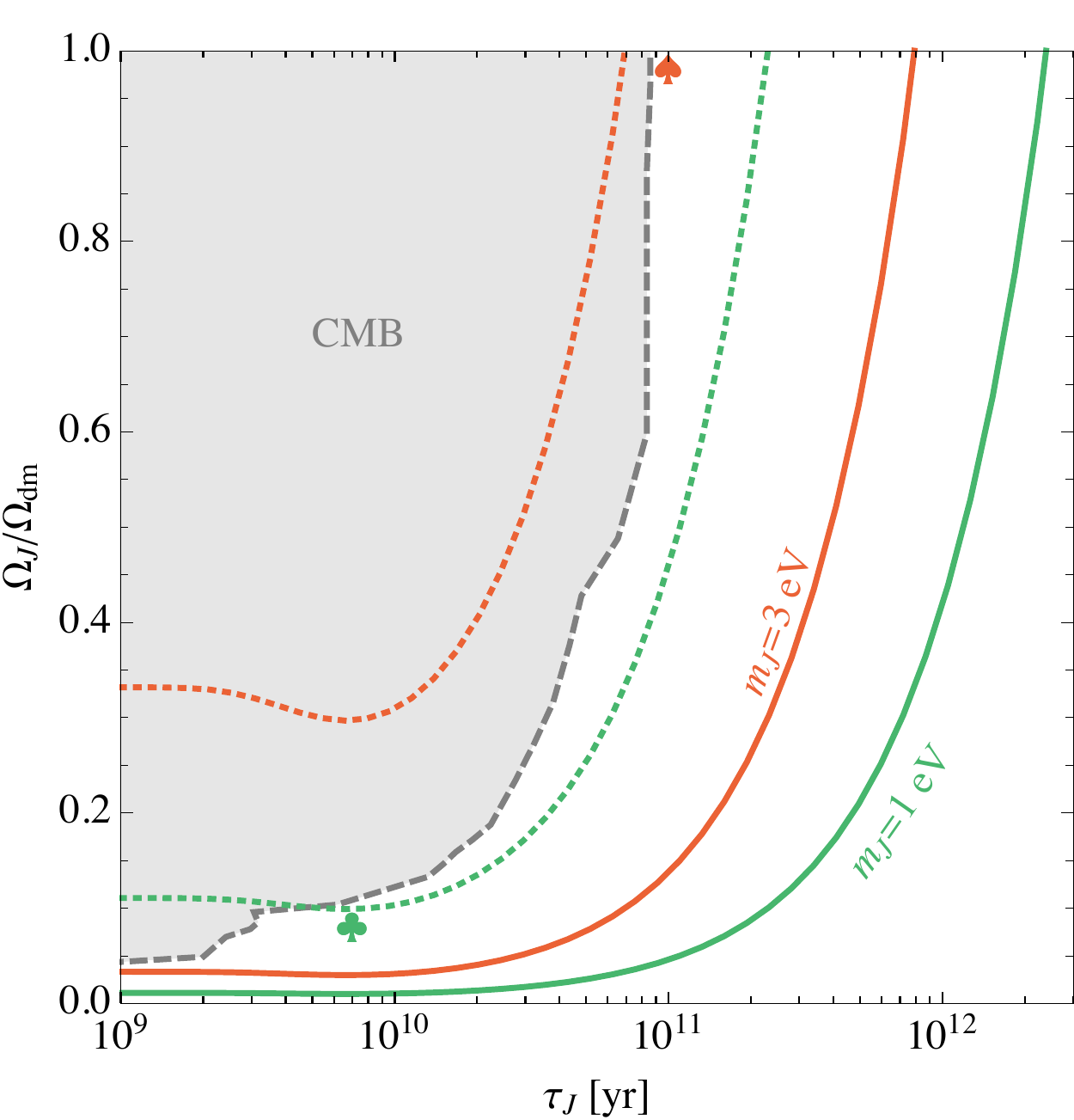}
\end{center}
\caption{Contours showing capture rates on tritium of $1~{\rm yr}^{-1}(100~\rm g)^{-1}$ (solid) and $10~{\rm yr}^{-1}(100~\rm g)^{-1}$ (dotted) in the space of $\tau_J$ and the $J$ energy density in units of the total dark matter energy density, $\Omega_J/\Omega_{\rm dm}$ ($J$ acts as a decaying dark matter component). $m_J$ has been chosen to be, from right to left, 1 (green) and 3 (red) eV. In all cases we have taken the electron neutrino fraction of the $J$ decays to be $f_{\nu_e}=1/2$. The $2\sigma$ exclusion from CMB data in~\cite{Poulin:2016nat} is the shaded gray region. We also show the locations of the two benchmark points which are plotted in Figs.~\ref{fig:flux} and \ref{fig:rate}: $m_J=1~\rm eV$, $\tau_J=7~\rm Gyr$, $\Omega_J/\Omega_{\rm dm}=0.08$ (green club) and $m_J=3~\rm eV$, $\tau_J=10^{11}~\rm yr$, $\Omega_J/\Omega_{\rm dm}=1$ (red spade).}\label{fig:reach}
\end{figure}

In Fig.~\ref{fig:reach}, we show contours for a capture rate on tritium of  $1~{\rm yr}^{-1}(100~\rm g)^{-1}$ [$10~{\rm yr}^{-1}(100~\rm g)^{-1}$] as solid [dotted] curves in the $\Omega_J/\Omega_{\rm dm}$ vs. $\tau_J$ parameter space, for (from right to left) $m_J=1,3~{\rm eV}$. We have chosen $f_{\nu_e}=1/2$, appropriate if $J$ were a Jajoron decaying to light neutrinos with an inverted mass hierarchy. Note that these are raw signal rates and do not impose a cut to remove background from $\beta$ decay. In addition, we shade the $2\sigma$ exclusion region on a decaying dark matter component from the analysis of CMB data in Ref.~\cite{Poulin:2016nat}. We also point out the two benchmark points in this parameter space whose fluxes were shown in Fig.~\ref{fig:flux}: $m_J=1~\rm eV$, $\tau_J=7~\rm Gyr$, $\Omega_J/\Omega_{\rm dm}=0.08$ as a green club and $m_J=3~\rm eV$, $\tau_J=10^{11}~\rm yr$, $\Omega_J/\Omega_{\rm dm}=1$ as a red spade.

\subsection{Signature}
Just as is the case with the C$\nu$B, detecting the neutrinos from $J$ decay is challenging because the signal of an electron above the $\beta$-decay end point can be swamped by the enormous rate from $\beta$ decay due to the finite $e^-$ energy resolution. To mitigate this background, PTOLEMY aims for roughly $0.1~\rm eV$ resolution on the electron energy to be able to observe the C$\nu$B for neutrinos with $m_\nu\gtrsim 0.1~\rm eV$. For reference, Dirac (Majorana) C$\nu$B neutrinos would lead to a rate of $4~(8)~\rm yr^{-1}$ for $100~\rm g$ of tritium~\cite{Long:2014zva}. We see that $J\to\nu\nu$ decays at late times can lead to comparable event rates with $E_\nu$ above threshold and, unlike the C$\nu$B, this can even be the case if $m_\nu<0.1~\rm eV$.

\begin{figure}
\begin{center}
\includegraphics[width=1.0\linewidth]{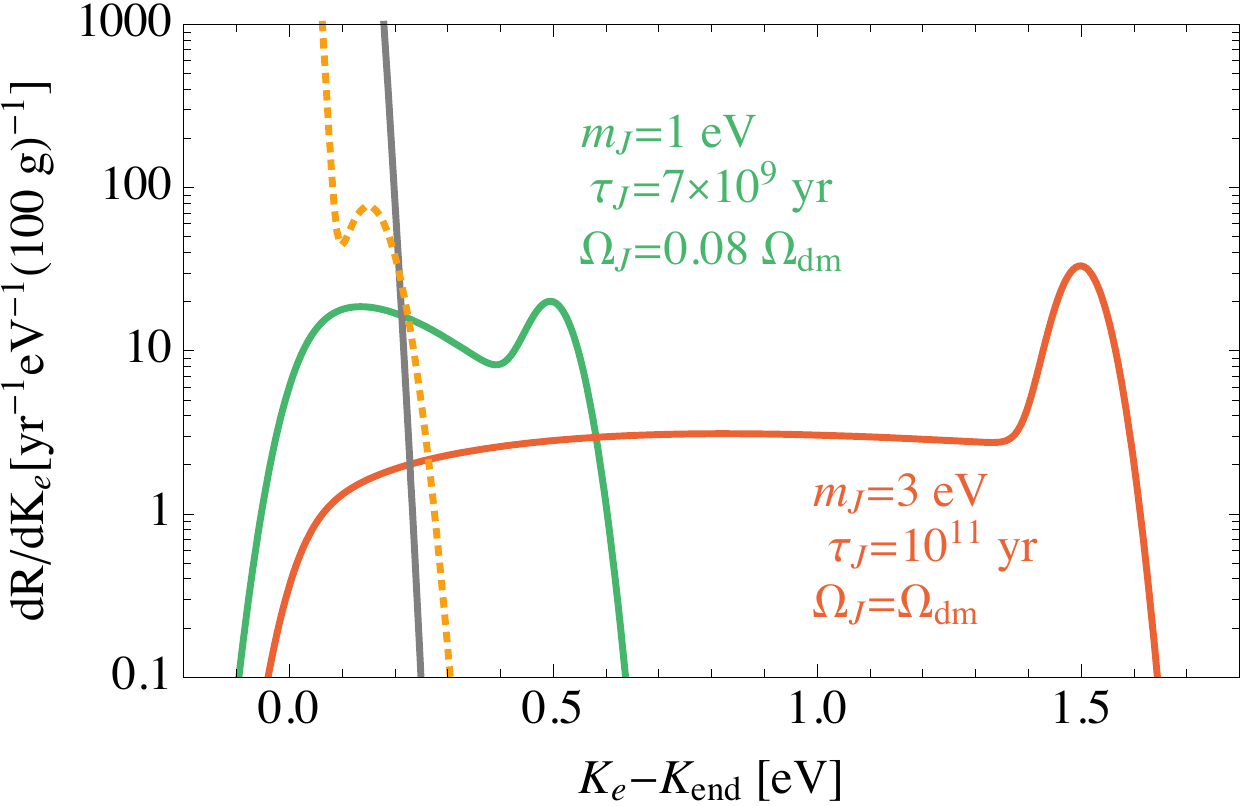}
\end{center}
\caption{Event rates for capture on tritium for neutrinos from $J\to\nu\nu$ for $m_J=1~\rm eV$, $\tau_J=7\times 10^{9}~\rm yr$, $\Omega_J/\Omega_{\rm dm}=0.08$ (green, solid) and $m_J=3~\rm eV$, $\tau_J=10^{11}~\rm yr$, $\Omega_J/\Omega_{\rm dm}=1$ (red, solid) and natural $\beta$ decay (gray, solid) with an electron energy resolution of $0.1~\rm eV$. For comparison, we plot the total rate for signal plus background for a Majorana C$\nu$B signal with $m_\nu=0.15~\rm eV$ (yellow, dashed).}\label{fig:rate}
\end{figure}

To illustrate the signal, we show the $e^-$ kinetic energy distribution for capture on tritium in Fig.~\ref{fig:rate}, assuming a resolution of $0.1~\rm eV$ as proposed by PTOLEMY, for two benchmark points currently allowed by cosmological data, $m_J=1~\rm eV$, $\tau_J=7\times10^{9}~\rm yr$, $\Omega_J/\Omega_{\rm dm}=0.08$ (solid, green) and $m_J=3~\rm eV$, $\tau_J=10^{11}~\rm yr$, $\Omega_J/\Omega_{\rm dm}=1$ (solid, red) taking $f_{\nu_e}=1/2$ as above. These benchmark points are also marked in Fig.~\ref{fig:reach} and their cosmically averaged fluxes are shown in Fig.~\ref{fig:flux}. Also shown in Fig.~\ref{fig:rate} is the background from the standard tritium $\beta$ decay with $m_\nu=0$ as a solid gray line. The rates are shown as functions of the $e^-$ kinetic energy $K_e$ minus the $m_\nu=0$ end point energy $K_{\rm end}$. The $m_J=1$ and $3~\rm eV$ benchmarks here give rates of $4.8$ and $7.0~{\rm yr}^{-1}(100~\rm g)^{-1}$ for $K_e-K_{\rm end}>0.2~\rm eV$, respectively, compared to $0.66~{\rm yr}^{-1}(100~\rm g)^{-1}$ from the $\beta$-decay background. As a comparison, we also plot the total rate in the case of a standard C$\nu$B signal with an $m_\nu=0.15~\rm eV$ Majorana neutrino as a dashed yellow curve.

Note that, for the diffuse components of the signals, we used, as in Fig.~\ref{fig:flux}, the energy spectrum from Eq.~(\ref{eq:spectrum}), $\Omega_{\rm b}=0.05$, $\Omega_\Lambda=0.69$, $\Omega_{\rm dm}=0.26$ with the dark matter split up into a decaying and nondecaying component, $\Omega_J$ and $\Omega_{\rm dm}-\Omega_J$, respectively. The contribution from $J$ decays within the Milky Way gives bumps with a width determined by the detector resolution. These two sources of neutrinos lead to the interesting signature of a peak at $K_e-K_{\rm end}=m_J/2$ from local $J$ decays and a shoulder extending from $m_J/2$ down to the end point from redshifted diffuse decays, with roughly comparable signal strength in each. The size of the signal from local $J$ decays is subject to uncertainty coming from our ignorance of the precise dark matter distribution of the Milky Way as well as its overall normalization; observation of a nonzero signal could of course help shed light on this issue.

\section{Conclusions}
\label{sec:conc}
This paper has explored a new physics scenario that searches for the C$\nu$B can impact: light dark matter that decays to neutrinos. While the event rates at a PTOLEMY-like experiment in this scenario are not enormous and the range of $m_J$ accessible is not large, over a spread of several eV, a comparable signal strength to those expected from the C$\nu$B is possible. Furthermore, the signal described here is distinct from that of the C$\nu$B, and can potentially be observable even if the $e^-$ energy resolution is poorer than what is needed for C$\nu$B detection given the current upper bound on $m_\nu$. 

While a robust statistical analysis of possible signals is beyond the scope of this study, if the electron energy resolution can be kept to the roughly $0.1~\rm eV$ level, excesses beyond the $\beta$-decay background could potentially be seen. Seeing such a signal would of course be tremendously exciting, opening up a world of new physics implications. Even providing an upper bound on the rate, however, would constrain part of the parameter space of dark matter decaying to neutrinos that is not currently probed by cosmological observations. This would be useful information, especially in light of nagging tensions in the cosmological data, such as the discrepancy in the extraction of the Hubble constant between CMB and local observations and the $\sigma_8$ problem. This signature also does not suffer from the same degeneracies that affect the extraction of conclusions from cosmological observations. Furthermore, in the event of a positive signal, the two distinct sources of neutrinos, from $J$ decays averaged over the universe and those in the Milky Way could allow for the dark matter distribution of the Milky Way to be studied in more detail. Another interesting possibility in the event of a positive signal would come from terrestrial searches for light particles coupled to neutrinos, e.g. imprinting spectral features in the spectrum of tritium $\beta$ decay~\cite{Arcadi:2018xdd}.

Similar models with a higher multiplicity of neutrinos in the decay of dark matter $\chi$, e.g. $\chi\to 2J\to4\nu$, could lead to larger event rates. However, this would come at the cost of diluting the energy of the neutrinos today, potentially masking the distinguishing spectral features in the capture rate. Other models, more complicated than the simple one described here, with potentially more involved cosmological histories could also lead to larger enhancements. 

Although we have not discussed them in detail, other proposals to search for the C$\nu$B, such as using laser interferometers~\cite{Domcke:2017aqj}, that do not involve looking for electrons above the $\beta$ decay endpoint could also be impacted by the decay of dark matter to neutrinos. While potential signals in such a setup are not likely to lead to large effects, assessing the sensitivity of new techniques in general to the signal described in this paper would be interesting. Moreover, other interesting features to consider would be the effects of gravitational clustering of the neutrinos from $J$ decay and potential anisotropies in the signal from decays in the Milky Way, along the lines explored in~\cite{Lisanti:2014pqa}.

\begin{acknowledgments}
We thank Nikita Blinov, Andrew Long, David Morrissey, Ann Nelson, Maxim Pospelov, and Nirmal Raj for helpful conversations. This work is supported by the National Research Council of Canada and was performed in part at the Aspen Center for Physics, which is supported by National Science Foundation grant PHY-1607611.
\end{acknowledgments}

\bibliography{latenus}

\end{document}